\documentclass[10pt,journal,a4paper]{IEEEtran} 

\usepackage{amssymb}
\usepackage{amsmath}
\usepackage{amsmath,bm}
\usepackage{amsthm}
\usepackage{graphicx}
\usepackage{subfigure}
\usepackage{cite}
\usepackage{enumerate}
\usepackage{algorithm}
\usepackage{algorithmicx}  
\usepackage{algpseudocode} 
\usepackage{color, soul}
\usepackage{stfloats}
\allowdisplaybreaks[4]
\begin{document}

\newtheorem{definition}{\bf ~~Definition}
\newtheorem{observation}{\bf ~~Observation}
\newtheorem{theorem}{\bf ~~Theorem}
\newtheorem{proposition}{\bf ~~Proposition}
\newtheorem{remark}{\bf ~~Remark}

\renewcommand{\algorithmicrequire}{\textbf{Input:}} 
\renewcommand{\algorithmicensure}{\textbf{Output:}} 

\title{\Large{Reconfigurable Intelligent Surfaces assisted Communications with Limited Phase Shifts: \\How Many Phase Shifts Are Enough?}}
\author{
{Hongliang Zhang}, \IEEEmembership{Member, IEEE},
{Boya Di}, \IEEEmembership{Member, IEEE},\\
{Lingyang Song}, \IEEEmembership{Fellow, IEEE},
{and Zhu Han}, \IEEEmembership{Fellow, IEEE}

\thanks{H. Zhang is with Department of Electronics, Peking University, Beijing, China, and also with Electrical and Computer Engineering Department, University of Houston, Houston, TX, USA (Email: hongliang.zhang92@gmail.com).}

\thanks{B. Di is with Department of Electronics, Peking University, Beijing, China, and also with Department of Computing, Imperial College of London, London, UK  (Email: diboya92@gmail.com).}

\thanks{L. Song is with Department of Electronics, Peking University, Beijing, China  (Email: lingyang.song@pku.edu.cn).}


\thanks{Z. Han is with Electrical and Computer Engineering Department, University of Houston, Houston, TX, USA, and also with the Department of Computer Science and Engineering, Kyung Hee University, Seoul, South Korea (Email: zhuhan22@gmail.com).}
\vspace{-5mm}}

\maketitle

\begin{abstract}
Reconfigurable intelligent surface~(RIS) has drawn a great attention worldwide as it can create favorable propagation conditions by controlling the phase shifts of the reflected signals at the surface to enhance the communication quality. However, the practical RIS only has limited phase shifts, which will lead to the performance degradation. In this paper, we evaluate the performance of an uplink RIS assisted communication system by giving an approximation of the achievable data rate, and investigate the effect of limited phase shifts on the data rate. In particular, we derive the required number of phase shifts under a data rate degradation constraint. Numerical results verify our analysis.
\end{abstract}

\begin{keywords}
Reconfigurable intelligent surfaces assisted communications, limited phase shifts, performance analysis 
\end{keywords}

\section{Introduction}
Recently, reconfigurable intelligent surface~(RIS) has been proposed as a novel and cost-effective solution to achieve high spectral and energy efficiency for wireless communications via only low-cost reflecting elements~\cite{M-2019}. With a large number of elements whose electromagnetic response (e.g. phase shifts) can be controlled by simple programmable PIN diodes~\cite{TDR-2010}, the RIS can reflect the incident signal and generate a directional beam, and thus enhancing the link quality and coverage.

In literature, some initial works have studied the phase shifts optimizations in the RIS assisted wireless communications. Since the phase shifts on the RIS will significantly influence the received energy, it provides another dimension to optimize for further quality-of-services~(QoS) improvement. In \cite{XDR-2020}, beamforming and continuous phase shifts of the RIS were optimized jointly to maximize the sum rate for an RIS assisted point-to-point communication system. For multi-user cases, the authors in \cite{BHLYZH}, proposed a hybrid beamforming scheme for a multi-user RIS assisted multi-input multi-output (MIMO) system together with a limited phase shifts optimization algorithm to maximize the user data rate. In \cite{CAGMC-2019}, a joint power allocation and continuous phase shift design is developed to maximize the system energy efficiency.

However, most works assume continuous phase shift, which are hard to be implemented in practical systems~\cite{TDR-2010}. Therefore, it is worthwhile to study the impact of the limited phase shifts on the achievable data rate. In this paper, we consider an uplink cellular network where the direct link between the base station (BS) and the user suffer from deep fading. To improve the QoS at the BS, we utilize a practical RIS with limited phase shifts to reflect signal from the user to the BS. To evaluate the performance limits of the RIS assisted communications, we provide an analysis on the achievable data rate with continuous phase shifts of the RIS, and then discuss how the limited phase shifts influence the data rate based on the derived achievable data rate.  

The rest of this paper is organized as follows. In Section \ref{System}, we introduce the system model for the RIS assisted communications. In Section \ref{Rate}, the achievable data rate is derived. The impact of limited phase shifts is discussed in Section \ref{analysis}. Numerical results in Section \ref{simulation} validate our analysis. Finally, conclusions are drawn in Section \ref{sec:conclusion}.

\section{System Model}%
\label{System}

Consider a narrow-band uplink cellular network\footnote{In the downlink case, although the working frequency and transmission power might be different, a similar method can be adopted since the channel model is the same due to the channel reciprocity.} consisting of one BS and one cellular user. Due to the dynamic wireless environment involving unexpected fading and potential obstacles, the Light-of-Sight~(LoS) link between the cellular user and the BS may not be stable or even falls into a complete outage. To tackle this problem, we adopt an RIS to reflect the signal from the cellular user towards the BS  to enhance the QoS. In the following, we first introduce the RIS assisted communication model, and then present the reflection-dominant channel model.

\begin{figure}[!t]
	\centering
	\includegraphics[width=2.8in]{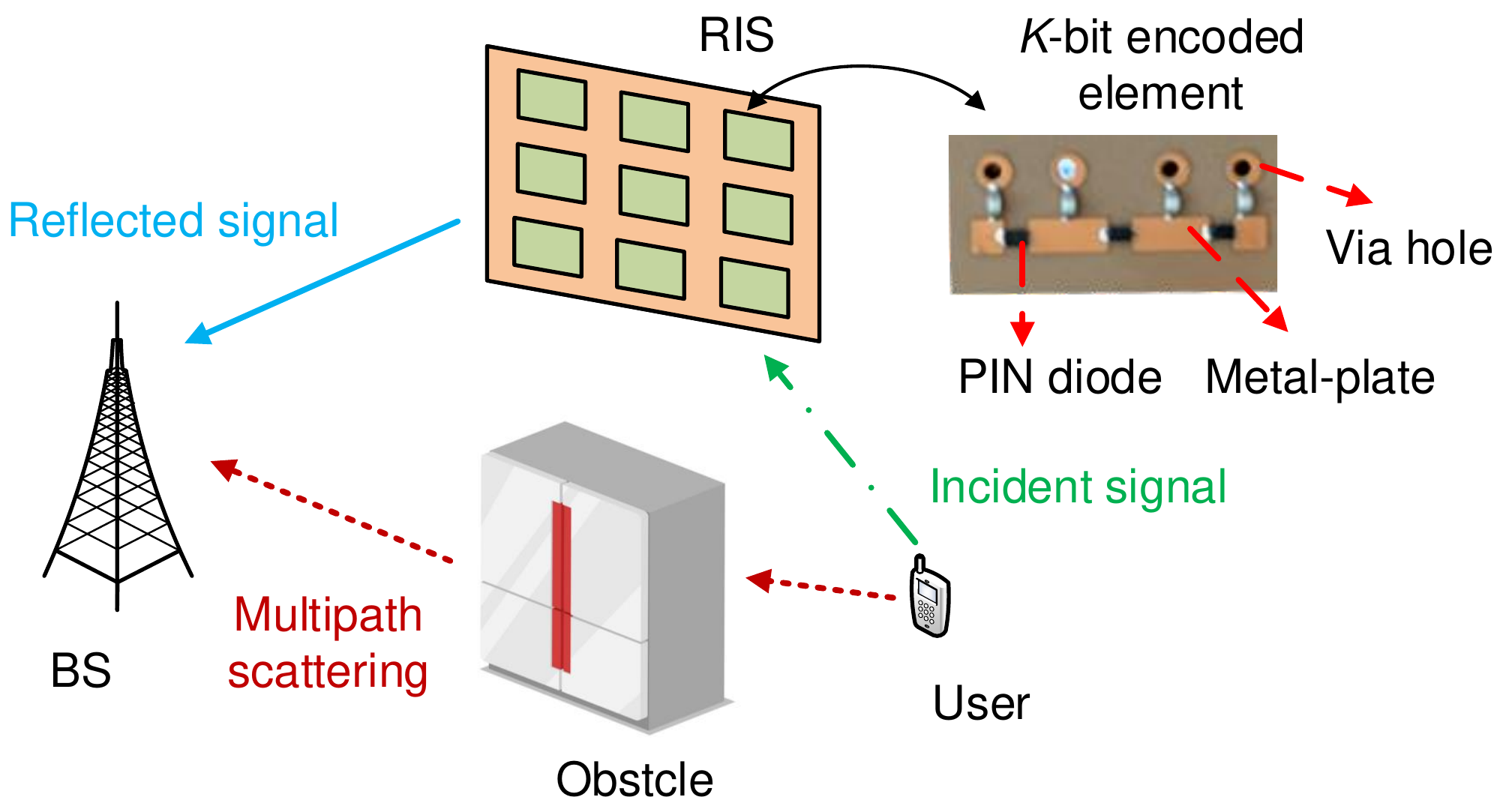}
		\vspace{-2mm}
	\caption{System model for the RIS assisted uplink cellular network.}
	\vspace{-4mm}
	\label{scenario}
\end{figure}

\subsection{RIS assisted Communication Model}
The RIS is composed of $M \times N$ electrically controlled RIS elements. Each element can adjust the phase shift by leveraging positive-intrinsic-negative~(PIN) diode. In Fig.~\ref{scenario}, we give an example of the element's structure. The PIN diode can be switched between ``ON" and ``OFF" states by controlling its biasing voltage, based on which the metal plate can add a different phase shift to the reflected signal. It is worthwhile to point out that the phase shifts are limited rather than continuous in practical systems~\cite{TDR-2010}. 

In this paper, we assume that the RIS is $K$ bit coded, that is, we can control the PIN diodes to generate $2^K$ patterns of phase shifts with a uniform interval $\Delta \theta = \frac{2\pi}{2^K}$. Therefore, the possible phase shift value can be given by $s_{m,n} \Delta \theta$, where $s_{m,n}$ is an integer satisfying $0 \leq s_{m,n} \leq 2^K - 1$. Without loss of generality, the reflection factor of RIS element $(m,n)$ at the $m$-th row and the $n$-th column is denoted by $\Gamma_{m,n}$, where
\begin{equation}\label{response}
\Gamma_{m,n} = \Gamma e^{-j \theta_{m,n}},
\end{equation}
where the reflection amplitude $\Gamma \in [0,1]$ is a constant.

\subsection{Reflection Dominant Channel Model}

\begin{figure}[!t]
	\centering
	\includegraphics[width=2.6in]{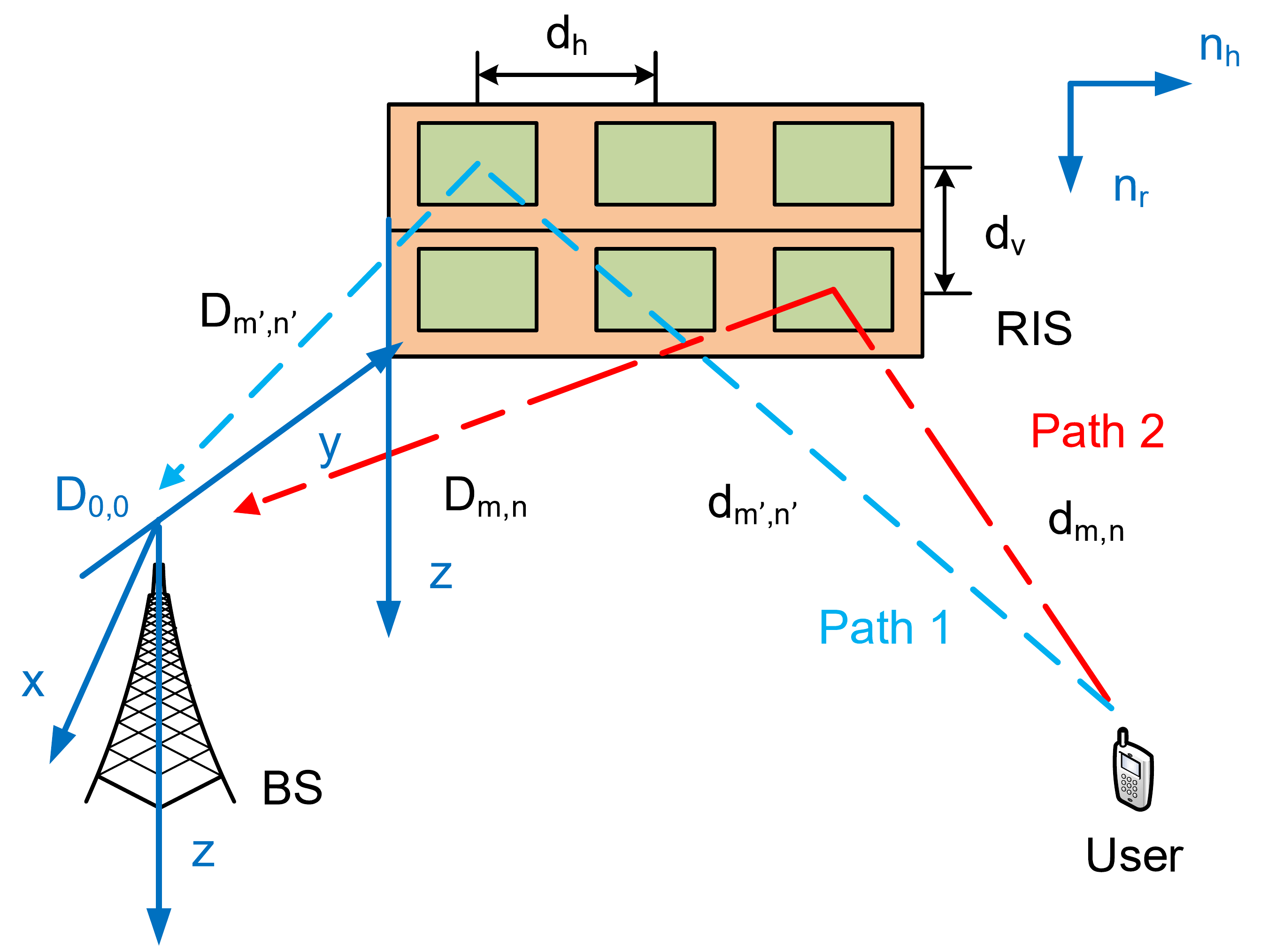}
	\vspace{-2mm}
	\caption{Channel model for the RIS-based uplink cellular network.}
	\vspace{-4mm}
	\label{Channel}
\end{figure}

In this subsection, we introduce the channel modeling between the BS and the user. Benefited from the directional reflections of the RIS, the received power from the BS-RIS-user links actually is much stronger than the multipath effect as well as the degraded direct link between the BS and the user. For this reason, we use the Rician model to model the channel. Here, the BS-RIS-user links act as the dominant LoS component and all the other paths contributes to the non-LoS~(NLoS) component. Therefore, the channel model between the BS and the user via RIS element $(m,n)$ can be written by
\begin{equation}
\tilde{h}_{m,n} = \sqrt{\frac{\kappa}{\kappa + 1}} h_{m,n} + \sqrt{\frac{1}{\kappa + 1}}\hat{h}_{m,n},
\end{equation}
where $h_{m,n}$ is the LOS component, $\hat{h}_{m,n}$ is the NLOS component, and $\kappa$ is the Rician factor indicating the ratio of the LoS component to the NLoS one.

As shown in Fig.~\ref{Channel}, let $D_{m,n}$ and $d_{m,n}$ be the distance between the BS and RIS element $(m,n)$, and that between element $(m,n)$ and the user, respectively. Define the transmission distance through element $(m,n)$ as $L_{m,n}$, where $L_{m,n} = D_{m,n} + d_{m,n}$. According to~\cite{A-2005}, the reflected LoS component of the channel between the BS and the user via RIS element $(m,n)$ can then be given by
\begin{equation}
\begin{array}{ll}
h_{m,n} & = \sqrt{G D_{m,n}^{-\alpha}d_{m,n}^{-\alpha}} e^{-j\frac{2 \pi}{\lambda} L_{m,n}}\\
& = \sqrt{G} \left[\sqrt{D_{m,n}^{-\alpha}}e^{-j\frac{2 \pi}{\lambda} D_{m,n}}\right]\cdot\left[\sqrt{d_{m,n}^{-\alpha}}e^{-j\frac{2 \pi}{\lambda} d_{m,n}}\right]\\
&= \sqrt{G}h^{t}_{m,n}h^r_{m,n},
\end{array}
\end{equation}
where $\alpha$ is the channel gain parameter, $G$ is the antenna gain, and $\lambda$ is the wave length of the signal. Here, $h_{m,n}^t$ and $h_{m,n}^r$ are the channel between the BS and RIS element $(m,n)$, as well as that between RIS element $(m,n)$ and the user, respectively. Similarly, the NLoS component can be written by
\begin{equation}
\hat{h}_{m,n} = PL(D_{m,n})PL(d_{m,n})g_{m,n},
\end{equation}
where $PL(\cdot)$ is the channel gain for the NLoS component and $g_{m,n} \sim \mathcal{CN}(0,1)$ denotes the small-scale NLoS components.

Using the geometry information, we can rewritten the channel gain by the following proposition, and the proof can be found in Appendix \ref{proof_appro}. 
\begin{proposition}\label{appro}
	When the distance $D_{m,n}$ between RIS element $(m,n)$ and the BS and the distance $d_{m,n}$ between element $(m,n)$ and the user  are much larger than the horizontal and vertical distances between two adjacent elements, $d_h$ and $d_v$, i.e., $D_{m,n}, d_{m,n} \gg d_h, d_v$, for $\forall m,n$, we have
	\begin{equation}\label{channel}
	G D_{m,n}^{-\alpha}d_{m,n}^{-\alpha} \triangleq PL_{LoS},~PL(D_{m,n})PL(d_{m,n}) \triangleq PL_{NLoS},
	\end{equation}
	where $PL_{LoS}$ and $PL_{NLoS}$ are constants.
\end{proposition}

Besides, we can have the following remark to show how the location of the RIS influences the channel gain.
\begin{remark} \label{remark1}
Given the transmission distance, i.e., $D_{m,n} + d_{m,n} = L$, where $L$ is a constant, the channel gain will decrease first and then increase when the RIS is further to the BS.	
\end{remark}
\begin{IEEEproof}
	For the LoS component, according to (\ref{channel}), we have
	\begin{equation}
	\begin{array}{ll}
	PL_{LoS} &= G((L - D_{m,n})D_{m,n})^{-\alpha} \\
	&= G(-(D_{m,n} - L/2)^2 + L^2/4)^{-\alpha}. 
	\end{array}
	\end{equation}
	Therefore, when $D_{m,n}$ increases, the LoS channel gain will decrease first and then increase. This trend will be the same for the NLoS component but the LoS component is typically dominant, and thus, the NLoS one can be neglected. This ends the proof.
\end{IEEEproof}


\section{Achievable Data Rate Analysis}\label{Rate}

After traveling through the reflection dominated channel, the received signal at the user can be expressed by
\begin{equation}
y = \sum\limits_{m,n} \Gamma_{m,n} \tilde{h}_{m,n} \sqrt{P} s + w,
\end{equation}
where $w \sim \mathcal{CN}(0,\sigma^2)$ is the additive white Gaussian noise, $P$ is the transmit power, and $s$ is the transmitted signal with $|s|^2 = 1$. Therefore, the received Signal-to-Noise Ratio (SNR) can be expressed by
\begin{equation}
\gamma = \frac{P}{\sigma^2} \left(\sum\limits_{m,n} \Gamma_{m,n} \tilde{h}_{m,n}\sum\limits_{{m'},{n'}}\Gamma^{*}_{{m'},{n'}}\tilde{h}^{*}_{m,n}\right),
\end{equation}
where $s^{*}$ is the conjugate of a complex number $s$.

\setcounter{equation}{17}
\begin{figure*}[hb]	
	\hrulefill
	\begin{equation}\label{required}
	K_{req} = \log_2 \pi - \log_2 \arccos \sqrt{\frac{\kappa + 1}{\kappa \eta_{LoS}M^2N^2}\left(\left(1 + \frac{\eta_{NLoS}}{\kappa + 1}MN + \frac{\kappa \eta_{LoS}}{\kappa + 1}M^2N^2\right)^{\epsilon_0} -1 - \frac{\eta_{NLoS}}{\kappa + 1}MN\right)},
	\end{equation}
\end{figure*}

The received SNR can be maximized by optimizing the response of each RIS element, and thus the achievable data rate can be expressed by
\setcounter{equation}{8}
\begin{equation}\label{achievable}
R = \max\limits_{\{\theta_{m,n}\}} \mathbb{E}\left[ \log_2(1 +\gamma)\right],
\end{equation}
where
\begin{equation}\label{datarate}
\begin{array}{ll}
&\hspace{-5mm}\mathbb{E}\left[ \log_2(1 +\gamma)\right] \approx \\
&~~\hspace{-8mm}\log_2 \left(1 \hspace{-1mm}+\hspace{-1mm} \frac{\eta_{LoS}}{\kappa + 1}MN\hspace{-1mm}+\hspace{-1mm}\frac{\kappa\eta_{NLoS}}{\kappa + 1}\hspace{-5mm}\sum\limits_{m,m',n, n'}\hspace{-5mm}e^{-j[\phi_{m,n} - \phi_{m',n'} + \theta_{m,n} - \theta_{m',n'}]}\right).
\end{array}
\end{equation}
Here, $\eta_{NLoS} = \frac{P\Gamma^2}{\sigma^2}PL_{NLOS}$, $\eta_{NLoS} =  \frac{P\Gamma^2}{\sigma^2} PL_{LOS}$, and $\phi_{m,n} = \frac{2 \pi}{\lambda}L_{m,n}$. Derivations are given in Appendix \ref{proof_datarate}.

To maximize the data rate, we need to let $\phi_{m,n} - \phi_{m',n'} + \theta_{m,n} - \theta_{m',n'} = 0$ for any $(m,n)$ and $(m',n')$. Thus, we have the following proposition.
\begin{proposition}\label{optimal}
The optimal phase shifts with continuous value $\theta_{m,n}^{*}$ should satisfy the following equation:
\begin{equation}\label{phase}
\theta_{m,n}^{*} + \phi_{m,n} = C,
\end{equation}
where $C$ is an arbitrary constant. The achievable data rate is
\begin{equation}\label{AD}
R = \log_2\left(1 + \frac{\eta_{NLoS}}{\kappa + 1}MN + \frac{\kappa \eta_{LoS} }{\kappa + 1}M^2N^2 \right).
\end{equation}
\end{proposition}

Based on the expression of the achievable data rate, we can have the following remarks to show the upper and lower bounds of the data rate.

\begin{remark}\label{R1}
	The upper bound of the data rate is achieved when we consider the pure LoS channel, i.e., $\kappa \rightarrow \infty$, where an asymptotic received power gain of $O(M^2N^2)$ can be obtained. 
\end{remark}

\begin{remark}\label{R2}
	The lower bound of the data rate is achieved when we consider the Rayleigh channel, i.e., $\kappa \rightarrow 0$, where an asymptotic squared received power gain of $O(MN)$ can be obtained. 
\end{remark}

These two Remarks show that the data rate grows with $\kappa$, since the received SNR increases from the order of $O(MN)$ to that of $O(M^2N^2)$.

\section{Analysis on the Number of Phase Shifts}
\label{analysis}

In this section, we will discuss the influence of limited phase shifts on the data rate. Since the number of phase shifts is finite in practice, we will select the one which is the closest to the optimal one $\theta_{m,n}^{*}$ as given in (\ref{phase}), and denote it by $\hat{\theta}_{m,n}$. Define the phase shift errors caused by limited phase shifts as
\begin{equation}
\delta_{m,n} = \theta_{m,n}^{*} - \hat{\theta}_{m,n}.
\end{equation} 
With $K$ coding bits, we have $- \frac{2\pi}{2^{K + 1}} \leq \delta_{m,n} < \frac{2\pi}{2^{K + 1}}$.

The SNR expectation $\hat{\gamma}$ with limited phase shifts can be written by
\begin{equation}\label{bound}
\begin{array}{ll}
\mathbb{E}[\hat{\gamma}]  & =  \frac{\eta_{NLoS}}{\kappa + 1}MN + \frac{\kappa \eta_{LoS}}{\kappa + 1} \hspace{-4mm}\sum\limits_{m,n,m',n'}\hspace{-4mm} e^{-j(C+\delta_{m,n} -C - \delta_{m'n'})}\\
& = \frac{\eta_{NLoS}}{\kappa + 1}MN + \frac{\kappa \eta_{LoS}}{\kappa + 1}\left|\sum\limits_{m,n} e^{-j \delta_{m,n}}\right|^2.
\end{array}
\end{equation}
Since $K \geq 1$, we have $\frac{2\pi}{2^{K+1}} \leq \frac{\pi}{2}$. Thus, the following inequality holds: 
\begin{equation}
\left|\sum\limits_{m,n} e^{-j \delta_{m,n}}\right|^2  \geq \left|MN e^{-j \frac{2\pi}{2^{K+1}}}\right|^2 = M^2N^2\cos^2\left(\frac{2\pi}{2^{K+1}}\right).
\end{equation}

To quantify the data rate degradation, we define the error $\epsilon$ brought by limited phase shifts as the ratio of the data rate with limited phase shifts to that with continuous ones.  To guarantee the system performance, $\epsilon$ should be larger than $\epsilon_0$ with $\epsilon_0 < 1$, i.e.,
\begin{equation}
\epsilon = \log_2(1 + \mathbb{E}[\hat{\gamma}])/\log_2(1 + \mathbb{E}[\gamma]) \geq \epsilon_0.
\end{equation}
Recalling (\ref{bound}), we can obtain the requirements on the coding bits as
\begin{equation}\label{requirement}
\frac{1 + \frac{\eta_{NLoS}}{\kappa + 1}MN + \frac{\kappa \eta_{LoS}}{\kappa + 1}M^2N^2\cos^2\left(\frac{2\pi}{2^{K+1}}\right)}{\left(1 + \frac{\eta_{NLoS}}{\kappa + 1}MN + \frac{\kappa \eta_{LoS}}{\kappa + 1}M^2N^2\right)^{\epsilon_0}} \geq 1.
\end{equation}
Therefore, the required number of coding bits can be written as in (\ref{required}).

In the following, we will provide a proposition to discuss impact of the RIS size on the required number of coding bits, i.e., the number of phase shifts.

\begin{proposition}\label{bit-size}
	Given the performance threshold $\epsilon_0$, the required coding bits is a decreasing function of the RIS size. Specially, when the RIS size is sufficiently large, i.e., $MN \rightarrow \infty$, 1 bit is sufficient to satisfy the performance threshold.
\end{proposition}
\begin{IEEEproof}
We first discuss how the number of phase shifts influences the required coding bits. Define the RIS size $x = MN \geq 1$, and 
\setcounter{equation}{18}
\begin{equation}
f(x) = \frac{\kappa + 1}{\kappa \eta_{LoS}x^2}\left(\left(1 \hspace{-1mm}+\hspace{-1mm} \frac{\eta_{NLoS}}{\kappa + 1}x \hspace{-1mm}+\hspace{-1mm} \frac{\kappa \eta_{LoS}}{\kappa + 1}x^2\right)^{\epsilon_0} \hspace{-2mm}-1 \hspace{-1mm}-\hspace{-1mm} \frac{\eta_{NLoS}}{\kappa + 1}x\right).	
\end{equation}
According to (\ref{required}), the required coding bits $K$ has the same monotonicity as $f(x)$. Therefore, we only need to investigate how $f(x)$ changes when $x$ increases. 

Let $a = \frac{\kappa\eta_{LoS}}{\kappa + 1}$ and $b = \frac{\eta_{NLoS}}{\kappa + 1}$. We have
\begin{equation}
f'(x) = \frac{1}{ax^{3}}\left(2 +\hspace{-1mm}bx \hspace{-1mm}+\hspace{-1mm} \frac{(\epsilon_0 \hspace{-1mm}- \hspace{-1mm}1)2ax^2 \hspace{-1mm}+\hspace{-1mm}(\epsilon_0 \hspace{-1mm}-\hspace{-1mm} 2) bx\hspace{-1mm} -\hspace{-1mm} 2}{\left(1 \hspace{-1mm}+\hspace{-1mm} bx \hspace{-1mm}+ \hspace{-1mm} ax^2\right)^{1 -\epsilon_0}}\right).
\end{equation}	
Note that $x \geq 1$ and $\epsilon_0 \leq 1$, we have
\begin{equation}
\begin{array}{ll}
f'(x) &\leq \frac{x^{-3}}{a}\left(2 \hspace{-1mm}+\hspace{-1mm}bx \hspace{-1mm}+\hspace{-1mm} \left((\epsilon_0 \hspace{-1mm}- \hspace{-1mm}1)2ax^2 \hspace{-1mm}+\hspace{-1mm}(\epsilon_0 \hspace{-1mm}-\hspace{-1mm} 2) bx\hspace{-1mm} -\hspace{-1mm} 2\right) \right)\\
& = \frac{x^{-3}}{a}\left((\epsilon_0 \hspace{-1mm}- \hspace{-1mm}1)2ax^2 \hspace{-1mm}+\hspace{-1mm}(\epsilon_0 \hspace{-1mm}-\hspace{-1mm} 1) bx\right) \leq 0.
\end{array}
\end{equation}
This implies that $f(x)$ decreases as $x$ grows, i.e., the required number of coding bits decreases as the RIS size grows.	

When the size of the RIS is sufficiently large, i.e., $MN \rightarrow \infty$, since $\epsilon < 1$, we have $\frac{1}{M^2N^2}\left(\left(1 + \frac{\eta_{NLoS}}{\kappa + 1}MN + \frac{\kappa \eta_{LoS}}{\kappa + 1}M^2N^2\right)^{\epsilon_0} \hspace{-2mm}-1 - \frac{\eta_{NLoS}}{\kappa + 1}MN\right) \\= 0$. Therefore, the required number of coding bits should be $\log_2 \pi - \log_2 (\pi/2) = 1$.
\end{IEEEproof}

\section{Simulation Results}
\label{simulation}

\begin{figure}[!t]
	\centering
	\includegraphics[width=2.6in]{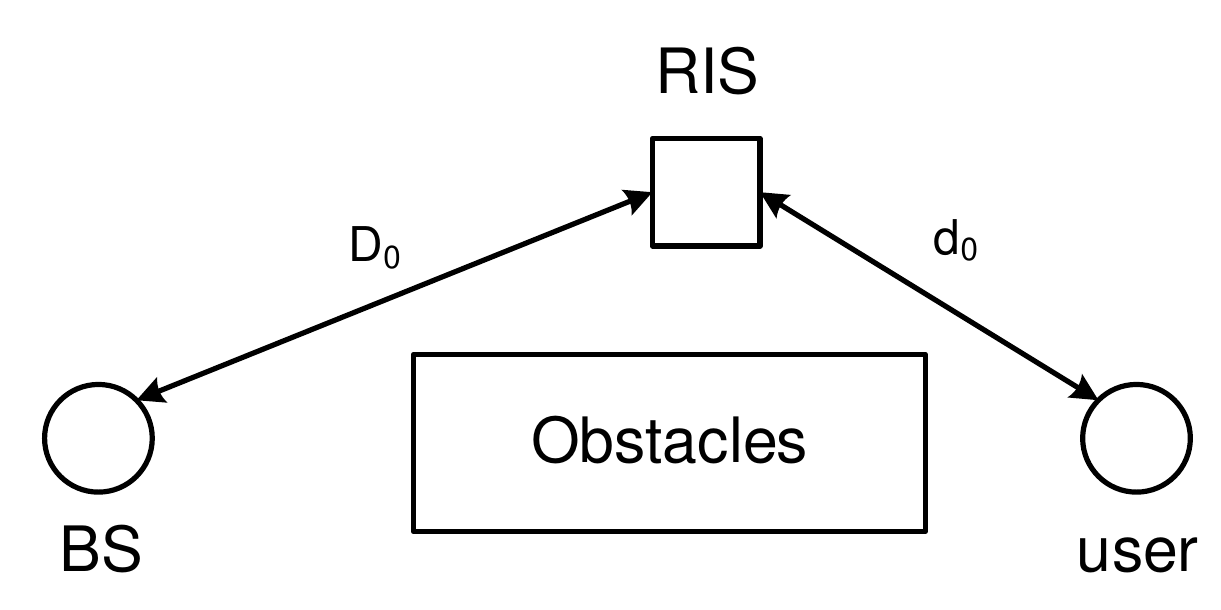}
	\vspace{-2mm}
	\caption{Simulation layout for the RIS-based cellular network (top view).}
	\vspace{-4mm}
	\label{layout}
\end{figure}

In this section, we verify the derivation of the achievable data rate and evaluate the optimal phase shift design for the RIS-assisted cellular network, the layout of which is given in Fig.~\ref{layout}. 
The parameters are selected according to 3GPP standard~\cite{3GPP-2018} and existing works \cite{TDR-2010,BHLYZH}.  The distances between the BS and the user to the center of the RIS are given by $D_0 = 95$ m and $d_0 = 65$ m, and the heights of the BS and the RIS are 25 m and 10 m, respectively. The RIS separation is set as $d_h = d_v = 0.03$ m and the reflection amplitude is assumed to be ideal, i.e., $\Gamma = 1$. The transmit power is $P = 20$ dBm, the noise power is $\sigma^2 = -96$ dBm, and the carrier frequency $f$ is 5.9 GHz. The UMa model in \cite{3GPP-2018} is utilized to describe the path loss for both LoS and NLoS components. For simplicity, we assume that $M = N$ in this simulation. All numeral results are obtained by 3000 Monte Carlo simulations.

\begin{figure}[!t]
	\centering
	\includegraphics[width=2.8in]{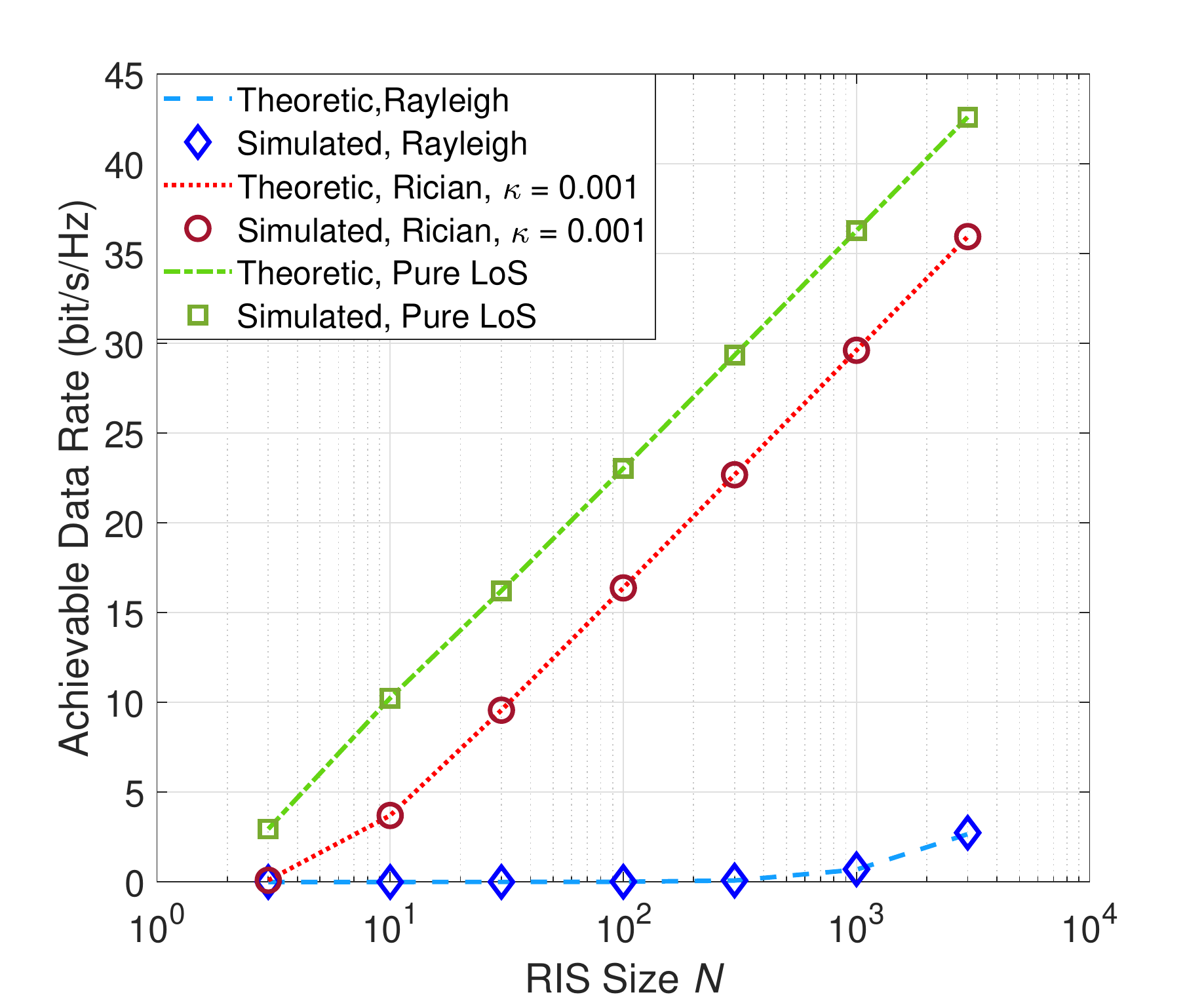}
	\vspace{-2mm}
	\caption{Achievable data rate vs. RIS size $N$ with continuous phase shifts.}
	\vspace{-4mm}
	\label{data}
\end{figure}

In Fig.~\ref{data}, we plot the achievable data rate vs. the RIS size $N$ with continuous phase shifts. From this figure, we can observe that our theoretic results are tightly close to the simulated ones. We can also observe that the data rate increases with the RIS size $N$, as more energy is reflected. In addition, when the RIS size is sufficiently large, the slope of the curve with the pure LoS channel is 4, which implies that the received SNR is proportional to the squared number of RIS elements\footnote{From (\ref{AD}), when $N$ is sufficiently large in the pure LoS scenario, the data rate can be written by $R = 4 \log_2(N) + z$, where $z$ is a constant. Since we use the logarithmic coordinates for the RIS size $N$, the slope of the curve being 4 is equal to the received SNR being an order of $O(N^4)$.}.  This result is consistent with Remark \ref{R1}. Similarly, when the RIS size is sufficiently large, the slope of the curve with the Rayleigh channel is 2, which corresponds to Remark \ref{R2}. In addition, we can observe that the data rate increases with the Rician factor, i.e., $\kappa$.

\begin{figure}[!t]
	\centering
	\includegraphics[width=2.8in]{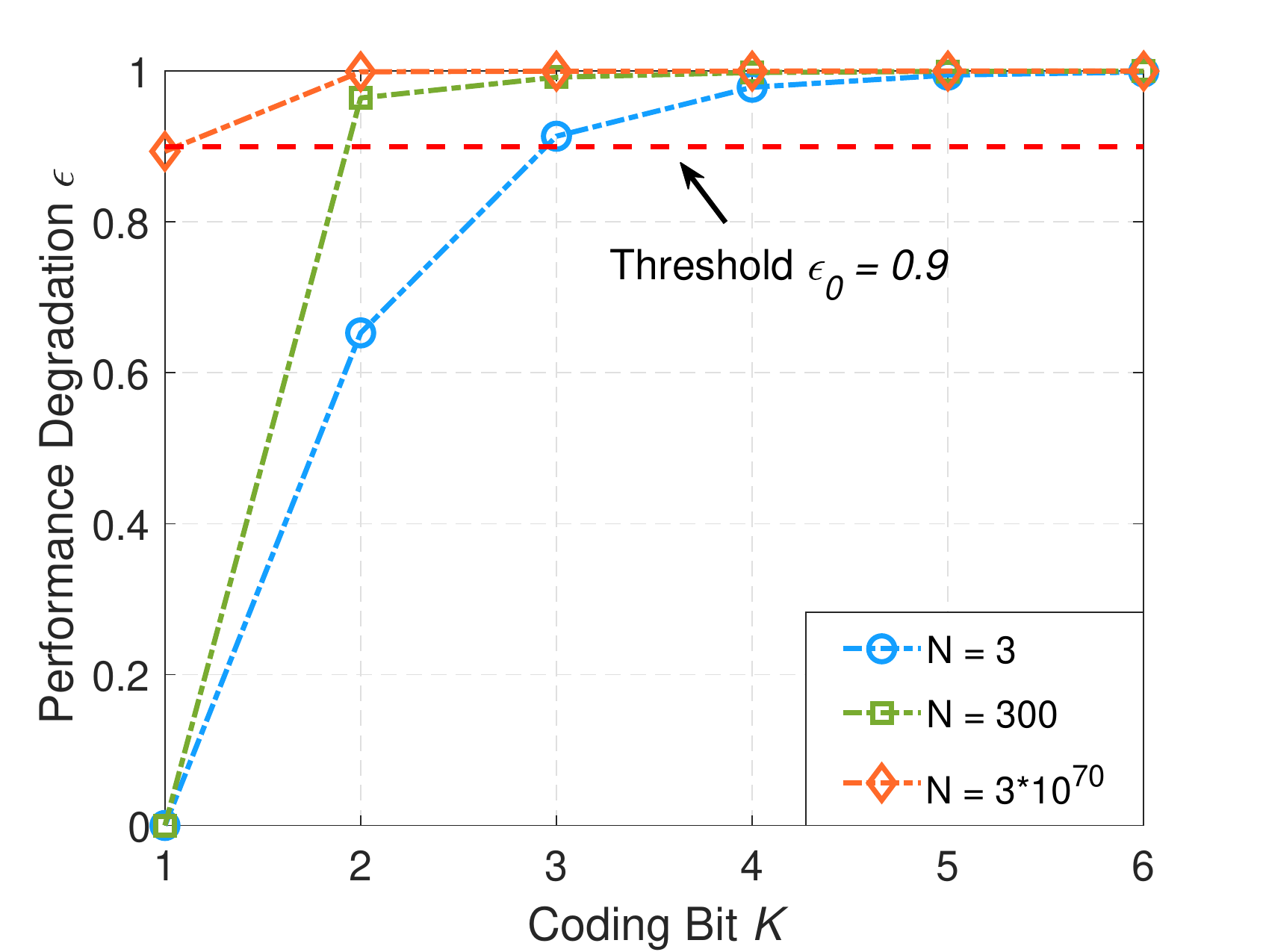}
	\vspace{-2mm}
	\caption{Data rate degradation $\epsilon$ vs. coding bit $K$ with $\kappa = 4$.}
	\vspace{-4mm}
	\label{quantization}
\end{figure}

In Fig.~\ref{quantization}, we plot the performance degradation $\epsilon$ vs. coding bit $K$ with $\kappa = 4$ in the Rician channel model \cite{BHLYZH} and threshold $\epsilon_0 = 0.9$. From this figure, we can find that the required number of phase shifts with the Rician channel model is: 1) 3 bit when the RIS size is small, e.g., $N = 3$; 2) 2 bit when the RIS size is moderate, e.g., $N = 300$; 3) 1 bit when the RIS size approaches to infinity, e.g., $N = 3*10^{70}$. These observations imply that the required coding bits decrease as the number of RIS elements grows, and 1 bit is enough when the RIS size goes to infinity, which verify Proposition \ref{bit-size}. 

\begin{figure}[!t]
	\centering
	\includegraphics[width=2.8in]{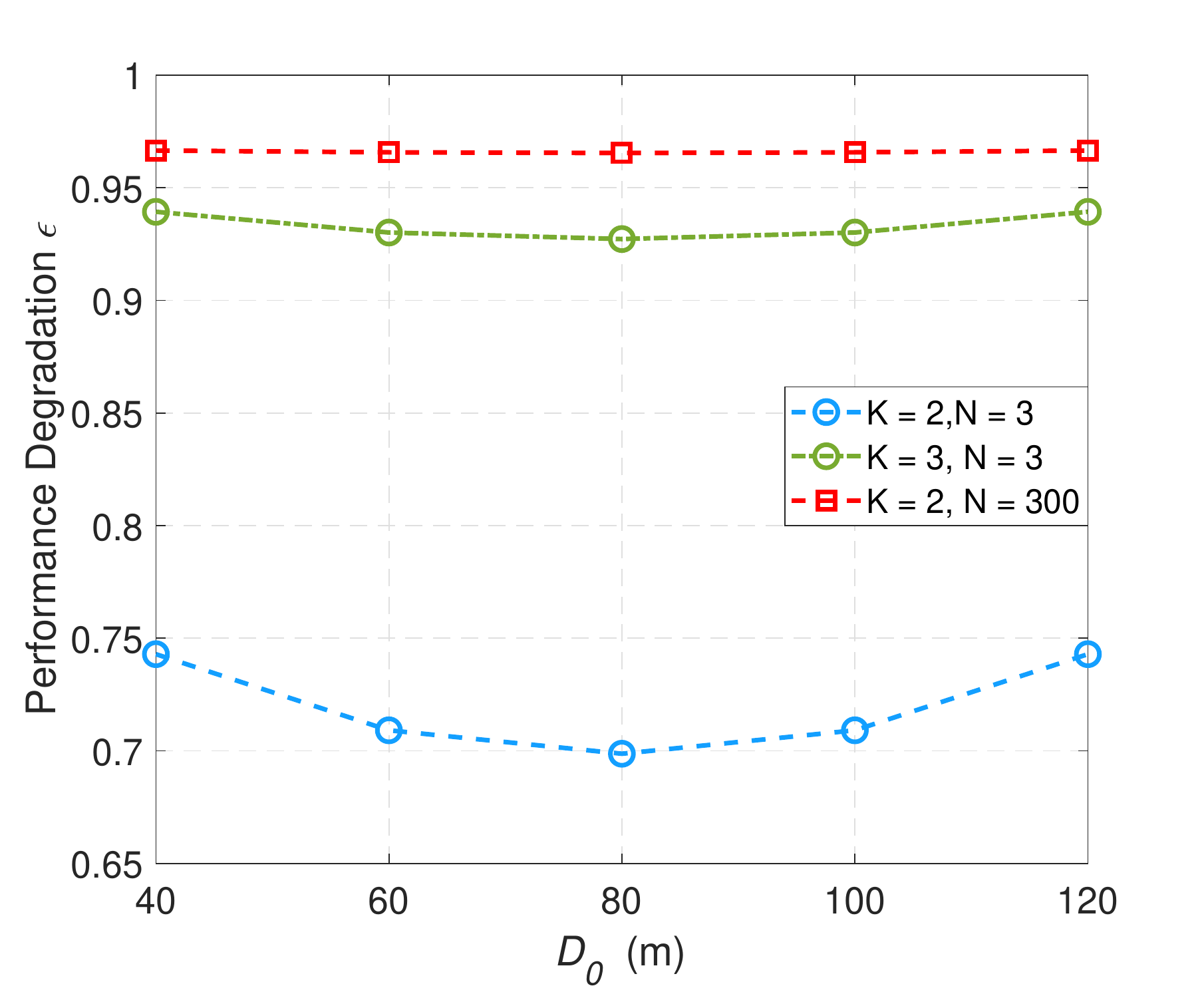}
	\vspace{-2mm}
	\caption{Data rate degradation $\epsilon$ vs. distance between the BS and the RIS $D_0$ with $\kappa = 4$ and $D_0 + d_0 = 160$.}
	\vspace{-4mm}
	\label{location}
\end{figure}

In Fig.~\ref{location}, we plot the performance degradation $\epsilon$ vs. distance between the BS and the RIS $D_0$ with $\kappa = 4$ to show how the location of the RIS influences the data rate degradation. For fairness, we assume that the transmission distance remains the same, i.e., $D_0 + d_0 = 160$. We can easily observe that the data rate degradation will decrease first and then increase as $D_0$ increases given RIS size $N$ and coding bit $K$. This is because the channel gain will decrease and then increase as $D_0$ increases, which has been proved in Remark \ref{remark1}, and the performance degradation $\epsilon$ is a increasing function of the channel gain\footnote{When coding bit $K$ and RIS size is given, the performance degradation $\epsilon$ can be written by $\epsilon = \frac{\log_2(1 + g'x)}{\log_2(1 + gx)}$, where $g' < g$ due to the limited phase shifts and $x$ is the channel gain. It is easy to check that this function is increasing as $x$ grows by calculating its first order derivative.}. In addition, we can also learn that the variance caused by the location change of the RIS will be smaller when the RIS size becomes larger. This implies that the location change of the RIS might influence the number of required coding bits when the RIS size is small, while the number of required coding bits keeps unchanged when the RIS size is sufficiently large.  

\vspace{-2mm}
\section{Conclusions and Future Works}
\label{sec:conclusion}
In this paper, we have derived the achievable data rate of the RIS assisted uplink cellular network and have discussed the impact of limited phase shifts design based on this expression. Particularly, we have proposed an optimal phase shift design scheme to maximize the data rate and have obtained the requirement on the coding bits to ensure that the data rate degradation is lower than the predefined threshold. 

From the analysis and simulation, given the location of the RIS, we can have the following conclusions: 1) We can achieve an asymptotic SNR of the squared number of RIS elements with a pure LoS channel model, and an asymptotic SNR of the number of RIS elements can be obtained when the channel is Rayleigh faded; 2) The required number of phase shifts will decrease as the RIS size grows given the data rate degradation threshold; 3) A number of phase shifts are necessary when the RIS size is small, while 2 phase shifts are enough when the RIS size is infinite. 

For future works, we can extend the RIS-assisted communications to Device-to-Device~(D2D) based heterogeneous network~\cite{ZMKGL-2016} for higher data rates since the interference among difference D2D users can be alleviated by proper phase shifts design, energy cooperation~\cite{HKM-2018} as the RIS can generate directional beams for more effective energy harvesting, and RF sensing~\cite{JHBLLYZH} for accurate object tracking by providing more information of the propagation environment.

\vspace{-2mm}
\begin{appendices}	
\section{Proof of Proposition \ref{appro}}\label{proof_appro}
Assume that the BS is located at the local origin and the angle between the direction of the RIS and the $x$-$y$ plane as $\theta_R$. Define the principle directions of the RIS as $\bm{n}_h$ and $\bm{n}_v$, we have $\bm{n}_h = \bm{n}_x \cos\theta_R + \bm{n}_y \sin\theta_R$ and $\bm{n}_v = \bm{n}_z$, where $\bm{n}_x$, $\bm{n}_y$, and $\bm{n}_z$ are directions of $x$, $y$, and $z$ axis.
	
Denote $\bm{c}_{m,n}$ as the position of RIS element $(m,n)$, where
\begin{equation}
\bm{c}_{m,n} = md_h\bm{n}_h + nd_v\bm{n}_v + D_{0,0}\bm{n}_y.
\end{equation}
Here, $D_{0,0}$ is the projected distance on $y$-asix between the BS and the RIS.
Therefore, we have
\begin{equation}
\begin{array}{ll}
D_{m,n}\hspace{-3mm} &= \left[(md_h\cos\theta_R)^2 + (md_h\sin\theta_R + D_{0,0})^2 +(nd_v)^2\right]^{\frac{1}{2}}\\
&  \approx (md_h\sin\theta_R + D_{0,0}) + \frac{(md_h\cos\theta_R)^2 +(nd_v)^2}{2 D_{0,0}},
\end{array}
\end{equation}
which can be achieved by $\sqrt{1 + a} \approx 1 + a/2$ when $a \ll 1$. We can obtain the expression of the distance between the RIS and the user using the similar method. Since the distance between two RIS elements is much smaller than that between the BS and user, the pathloss of the BS and user via different RIS element can be regarded a constant.	

\section{Derivations of Equation (\ref{datarate})}\label{proof_datarate}
Due to the property of the logarithmic function \cite{YWSCX-2019}, we have
\begin{equation}
\mathbb{E}[\log_2(1 + \gamma)] \approx \log_2(1 + \mathbb{E}[\gamma]).
\end{equation}
Since $\frac{P}{\sigma^2}$ is constant, we will derive $\mathbb{E}[\gamma]$ in the following.
\begin{equation}\label{expectation}
\begin{array}{ll}
\mathbb{E}[\gamma] \hspace{-3mm} & \hspace{-1mm}= \frac{P\Gamma^2}{\sigma^2} \mathbb{E}\left[\sum\limits_{m,n} e^{-j\theta_{m,n}} \tilde{h}_{m,n}\sum\limits_{{m'},{n'}}e^{j\theta_{m',n'}}\tilde{h}^{*}_{{m'},{n'}}\right]\\
&\hspace{-1mm}= \frac{P\Gamma^2}{\sigma^2}\hspace{-5mm}\sum\limits_{m,n,{m'},{n'}}\hspace{-5mm}e^{-j(\theta_{m,n}\hspace{-1mm} -\theta_{m',n'})}\left(\frac{1}{\kappa + 1} \mathbb{E}[\hat{h}_{m,n}\hat{h}^{*}_{{m'},{n}}] + \right.\\
&\hspace{-1mm}~\left.\frac{\kappa}{\kappa + 1} PL_{LoS} e^{-j[\phi_{m,n} - \phi_{{m'},{n'}}]} +\right. \\
&\hspace{-1mm}~\left.2\frac{\sqrt{\kappa}}{\kappa + 1}\sqrt{PL_{LOS}}\mbox{Re}\left\{ e^{j\phi_{m,n}}\mathbb{E}[\hat{h}_{m,n}]\right\} \right).
\end{array}
\end{equation}
Since $\hat{h}_{m,n}$ has a zero mean, the final term in (\ref{expectation}) equals to 0. Moreover, since $\hat{h}_{m,n}$ is independent for different elements $(m,n)$ and $(m',n')$, the following equation holds:
\begin{equation}
\mathbb{E}[\hat{h}_{m,n}\hat{h}^{*}_{m',n'}] = \left\{
\begin{array}{l}
PL_{NLoS}, \mbox{if}~m = m', n = n',\\
0, \mbox{otherwise}.
\end{array}
\right.
\end{equation}
Therefore, we have 
\begin{equation}
\vspace{-1mm}
\mathbb{E}[\gamma]=AMN+B\hspace{-4mm}\sum\limits_{m,m',n, n'}\hspace{-4mm}e^{-j[\phi_{m,n} - \phi_{m',n'} + \theta_{m,n} - \theta_{m',n'}]}.
\end{equation}
This ends the proof.
\end{appendices}

\end{document}